\pdfoutput=1

\documentclass[twocolumn,final]{svjour3}

\smartqed

\usepackage{graphicx,amssymb}
\usepackage{epsf,psfig}
\usepackage{txfonts}
\usepackage{natbib}

\journalname{Nonlinear Dynamics}

\begin{document}

\title{Exploring the origin, the nature and the dynamical behaviour of distant stars in galaxy models}

\author{Euaggelos E. Zotos}

\institute{Euaggelos E. Zotos:
\at Department of Physics, School of Science, \\
Aristotle University of Thessaloniki \\
GR-541 24, Thessaloniki, Greece \\
email:{evzotos@physics.auth.gr}
}

\date{Received: 3 June 2013 / Accepted: 7 July 2013 / Published online: 13 August 2013}

\titlerunning{Exploring the origin, the nature and the dynamical behaviour of distant stars in galaxy models}

\authorrunning{Euaggelos E. Zotos}

\maketitle

\begin{abstract}

We explore the regular or chaotic nature of orbits moving in the meridional plane of an axially symmetric galactic gravitational model with a disk, a dense spherical nucleus and some additional perturbing terms corresponding to influence from nearby galaxies. In order to obtain this we use the Smaller ALingment Index (SALI) technique integrating extensive samples of orbits. Of particular interest is the study of distant, remote stars moving in large galactocentric orbits. Our extensive numerical experiments indicate that the majority of the distant stars perform chaotic orbits. However, there are also distant stars displaying regular motion as well. Most distant stars are ejected into the galactic halo on approaching the dense and massive nucleus. We study the influence of some important parameters of the dynamical system, such as the mass of the nucleus and the angular momentum, by computing in each case the percentage of regular and chaotic orbits. A second order polynomial relationship connects the mass of the nucleus and the critical angular momentum of the distant star. Some heuristic semi-theoretical arguments to explain and justify the numerically derived outcomes are also given. Our numerical calculations suggest that the majority of distant stars spend their orbital time in the halo where it is easy to be observed. We present evidence that the main cause for driving stars to distant orbits is the presence of the dense nucleus combined with the perturbation caused by nearby galaxies. The origin of young O and B stars observed in the halo is also discussed.

\keywords{Galaxies: kinematics and dynamics; Numerical methods}

\end{abstract}

\section{Introduction}
\label{intro}

\begin{figure}
\includegraphics[width=\hsize]{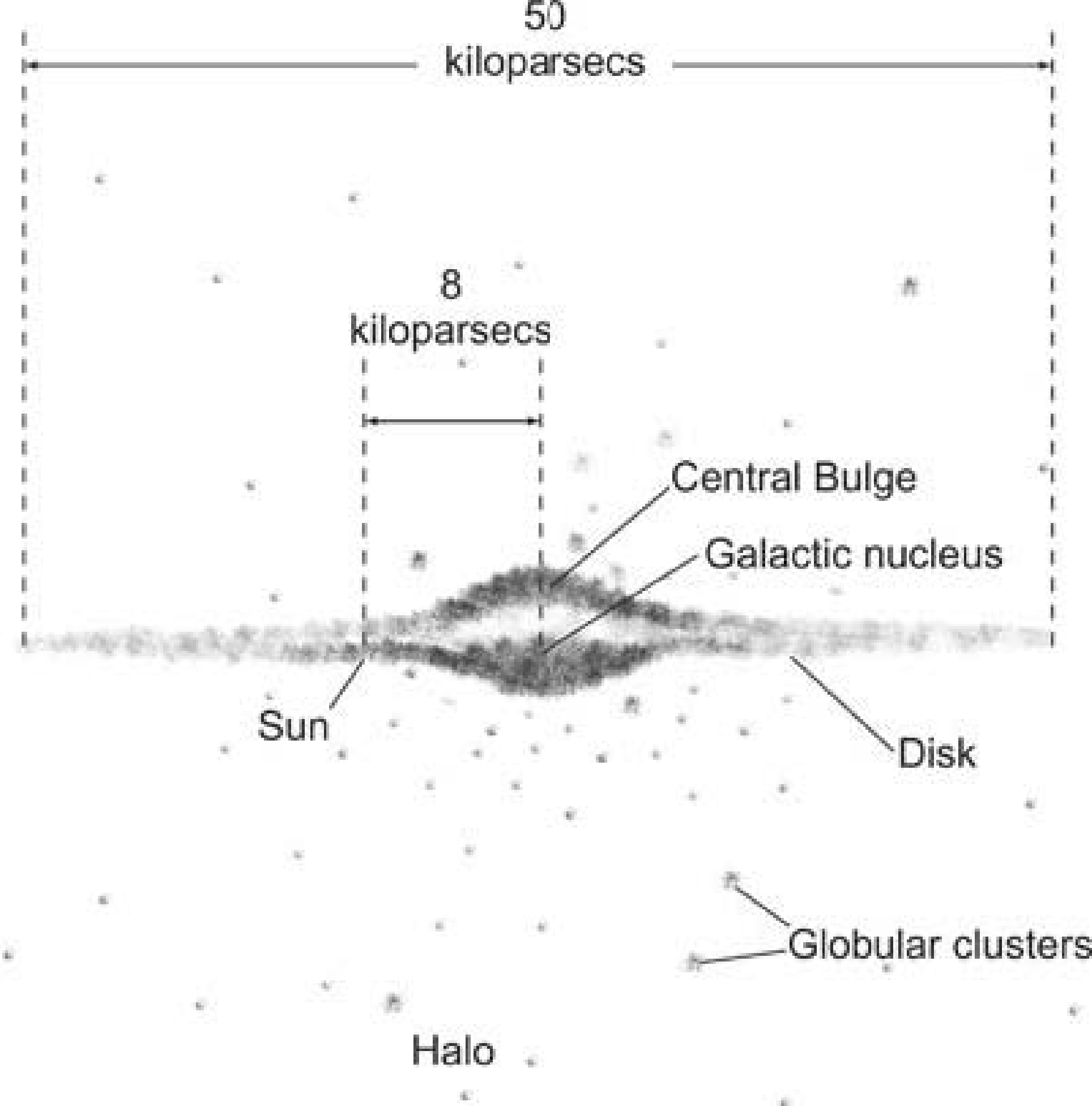}
\caption{An external view of the Milky Way galaxy, looking edge-on or sideways into the disk.}
\label{galstru}
\end{figure}

The Milky Way is believed to be a spiral galaxy, and the best ``educated guess" is that it is a barred Sb to Sc type of galaxy (e.g., [\citealp{4}, \citealp{19}]). However, since we are inside the Milky Way, it has been proved very difficult to properly characterize its structure [\citealp{18}]. While the greater part of the mass of the Milky Way lies in the relatively thin, circularly symmetric plane or disk, there are three other recognized components of the galaxy, each marked by distinct patterns of spatial distribution, motions and stellar types. These are the disk, the halo and the nucleus (see Fig. \ref{galstru}).

\vspace*{1\baselineskip}
\noindent \textbf{(1). The galactic disk:}

The disk consists of stars distributed in the thin, rotating, circularly symmetric plane that has an approximate diameter of 30 kpc and a thickness of about 400 to 500 pc. Most disk stars are relatively old, although the disk is also the site of present star formation as evidenced by the young open clusters and associations. The estimated present conversion rate of interstellar material to new stars is only about 1 solar mass ($\rm M_\odot$) per year. The Sun is a disk star about 8.5 kpc from the center of the galaxy. All these stars, old to young, are fairly homogeneous in their chemical composition, which is similar to that of the Sun.

The disk also contains essentially all the galaxy's content of interstellar material, but the gas and dust are concentrated to a much thinner thickness than the stars; half the interstellar material is within about 25 pc of the central plane. Within the interstellar material, denser regions contract to form new stars. In the local region of the disk, the position of young O and B stars, young open clusters, young Cepheid variables, and HII regions associated with recent star formation reveal that star formation does not occur randomly in the plane but in a spiral pattern analogous to the spiral arms found in other disk galaxies. The disk of the galaxy is in dynamical equilibrium, with the inward pull of gravity balanced by motion in circular orbits. The disk is fairly rapidly rotating with a uniform velocity about 220 km/s. Over most of the radial extent of the disk, this circular velocity is reasonably independent of the distance outward from the center of the galaxy.

\vspace*{1\baselineskip}
\noindent \textbf{(2). The galactic halo:}

Some stars and star clusters (globular clusters) form the halo component of the galaxy. They surround and interpenetrate the disk, and are thinly distributed in a more or less spherical (or spheroidal) shape symmetrically around the center of the Milky Way. The halo is traced out to about 50 kpc , but there is no sharp edge to the galaxy; the density of stars simply fades away until they are no longer detectable. The halo's greatest concentration is at its center, where the cumulative light of its stars becomes comparable to that of the disk stars. This region is called the (nuclear) bulge of the galaxy; its spatial distribution is somewhat more flattened than the whole halo. There is also evidence that the stars in the bulge have slightly greater abundances of heavy elements than stars at greater distances from the center of the galaxy.

The halo stars consist of old, faint, red main sequence stars or old, red giant stars, considered to be among the first stars to have formed in the galaxy. Their distribution in space and their extremely elongated orbits around the center of the galaxy suggest that they were formed during one of the galaxy's initial collapse phases. Forming before there had been significant thermonuclear processing of materials in the cores of stars, these stars came from interstellar matter with few heavy elements. As a result, they are metal poor. At the time of their formation, conditions also supported the formation of star clusters that had about $10^6$ $\rm M_\odot$ of material, the globular clusters. Today there exists no interstellar medium of any consequence in the halo and hence no current star formation there. The lack of dust in the halo means that this part of the galaxy is transparent, making observation of the rest of the universe possible.

Halo stars can easily be discovered by proper motion studies. In extreme cases, these stars have motions nearly radial to the center of the galaxy, hence at right angles to the circular motion of the Sun. Their net relative motion to the Sun therefore is large, and they are discovered as high-velocity stars, although their true space velocities are not necessarily great. Detailed study of the motions of distant halo stars and the globular clusters shows that the net rotation of the halo is small. Random motions of the halo stars prevent the halo from collapsing under the effect of the gravity of the whole galaxy.

\vspace*{1\baselineskip}
\noindent \textbf{(3). The galactic nucleus:}

The nucleus is considered to be a distinct component of the galaxy. It is not only the central region of the galaxy where the densest distribution of stars (about 5 $\times$ $10^4$ stars per cubic parsec compared to about 1 star per cubic parsec in the vicinity of the Sun) of both the halo and disk occurs, but it is also the site of violent and energetic activity. The very center of the galaxy harbors objects or phenomena that are not found elsewhere in the galaxy. This is evidenced by a high flux of infrared, radio, and extremely short wavelength gamma radiation coming from the center, a specific infrared source known as Sagittarius A. Infrared emissions in this region show that a high density of cooler stars exists there, in excess of what would be expected from extrapolating the normal distribution of halo and disk stars to the center.

The nucleus is also exceptionally bright in radio radiation produced by the interaction of high-velocity charged particles with a weak magnetic field (synchrotron radiation). Of greater significance is the variable emission of
gamma rays, particularly at an energy of 0.5 MeV. This gamma-ray emission line has only one source which is the mutual annihilation of electrons with anti-electrons, or positrons, the source of which in the center has yet to be identified. Theoretical attempts to explain these phenomena suggest a total mass involved of $10^6$-$10^7$ $\rm M_\odot$ in a region perhaps a few parsecs in diameter. This could be in the form of a single object, a massive black hole; similar massive objects appear to exist in the centers of other galaxies that exhibit active nuclei (AGNs). By the standards of such active galaxies, however, the nucleus of the Milky Way is a quiet place, although interpretations of the observed radiation suggest the existence of huge clouds of warm dust, rings of molecular gas, and other complex features.

Recent data derived from observations indicate that the vast majority of distant stars are located in the halo and they are old. Since these stars are among the oldest stars in the galaxy they offer great insight into the formation and also the early evolution of the Milky Way galaxy. Nevertheless, many scientists claim that young O and B stars have been also identified in the galactic halo [\citealp{14}]. Naturally the following question arises: Do we have a low level of star formation in the halo, or the observed young stars are in fact disk stars that have been ejected from the galactic plane? If so, what exactly mechanism has driven these young stars in large galactocentric orbits? One of the main targets of this research is to provide some possible answers to these interesting questions, continuing in much more detail the initial work presented in [\citealp{27}].

The present paper is organized as follows: In Section \ref{model} we present the structure and the properties of our galactic model. In Section \ref{cometh} we provide a brief description of the computational methods we used in order to explore the regular or chaotic nature of orbits. In the following Section, we investigate how the mass of the nucleus and the angular momentum influences the character of the orbits. Moreover in the same section, we search for the origin of young stars, observed in the galactic halo. In Section \ref{semitheor} we try to find a numerical relationship connecting the critical value of the angular momentum with the mass of the nucleus. Then, we present some heuristic semi-theoretical arguments, in order to support and explain the numerically obtained outcomes. We conclude with Section \ref{disc}, where the discussion and the conclusions of this research are presented.

\section{Presentation and properties of the galactic model}
\label{model}

The importance of galactocentric orbits in studies of the Milky Way is well known. The usual approach has been to assume a stationary and axially symmetric potential of the Milky Way with three main contributors: disk, nucleus and halo. In this way galactocentric orbits of many objects have been calculated [\citealp{1},\citealp{5},\citealp{15},\citealp{26}]. Projected on the so-called meridional plane these orbits show a variety of shapes. In most cases one finds box-like orbits, however there are also quite different orbits. The main objective of the present research work is to investigate the dynamical properties of the motion of stars in the meridional plane of an axially symmetric disk galaxy with an additional spherical nucleus. For this purpose, we use the cylindrical coordinates $(R, \phi, z)$, where $z$ is the axis of the symmetry.

The total potential $V(R,z)$ in our model is the sum of a disk-halo potential $V_{\rm d}$, a central spherical component $V_{\rm n}$ and some additional perturbation terms $V_{\rm p}$. The first one is a generalization of the Miyamoto-Nagai potential [\citealp{25}] (see also [\citealp{9}] and [\citealp{12}])
\begin{equation}
V_{\rm d}(R,z) = \frac{- G M_{\rm d}}{\sqrt{b^2 + R^2 + \left(\alpha + \sqrt{h^2 + z^2}\right)^2}}.
\label{diskpot}
\end{equation}
Here $G$ is the gravitational constant, $M_{\rm d}$ is the mass of the disk, $\alpha$ is the scale length of the disk, $h$ corresponds to the disk's scale height, while $b$ is the core radius of the disk-halo component. Moreover, in order to describe the spherically symmetric nucleus we use a Plummer potential
\begin{equation}
V_{\rm n}(R,z) = \frac{- G M_{\rm n}}{\sqrt{R^2 + z^2 + c_{\rm n}^2}},
\label{nucpot}
\end{equation}
where $M_{\rm n}$ and $c_{\rm n}$ are the mass and the scale length of the nucleus, respectively. This potential has been used several times in the past to model the central mass component of a galaxy (e.g., [\citealp{20},\citealp{21}]). At this point, we must point out that the nucleus is not intended to represent a black hole nor any other compact object; but a bulge. Therefore, all the relativistic phenomena that may occur in the central region of a galaxy are completely out of the scope of this research.

The potential containing the perturbation terms corresponding to influence and interactions from nearby galaxies is given by
\begin{equation}
V_{\rm p}(R,z) = k R^3 + \lambda \left(R^2 + \beta z^2\right)^2,
\label{perpot}
\end{equation}
where $k$, $\lambda$ and $\beta$ are parameters. Similar perturbing terms had been used in [\citealp{36}] in order to describe and model how nearby galaxies influence the dynamical behavior of stars. The following lines of arguments justify our choice: it is well known, that in galaxies possessing dense and massive nuclei [\citealp{7},\citealp{9},\citealp{10},\citealp{38}] low angular momentum stars, moving close to the galactic plane are scattered to much higher scale heights upon approaching the central nucleus thus displaying chaotic motion. In a previous work [\citealp{38}], we presented the properties of low angular momentum stars and also we provided useful correlations between the chaotic motion of stars and some important quantities of the dynamical system such as the angular momentum and the mass of the nucleus. The novelty in the current work is the study of the dynamical behavior of stars moving in ``distant orbits". Here we should clarify, that by the term distant orbits we refer to orbits of stars that reach large galactocentric distances, of order of about 30 kpc or more. In order to have a better estimation of the situation, we agree to call distant orbits all the orbits that obtain galactocentric distances larger than 10 kpc. This is the exact reason for introducing the additional perturbation terms in our model. We will see later on, that these terms are the basic factor which leads stars to distant orbits.

\begin{figure}
\includegraphics[width=\hsize]{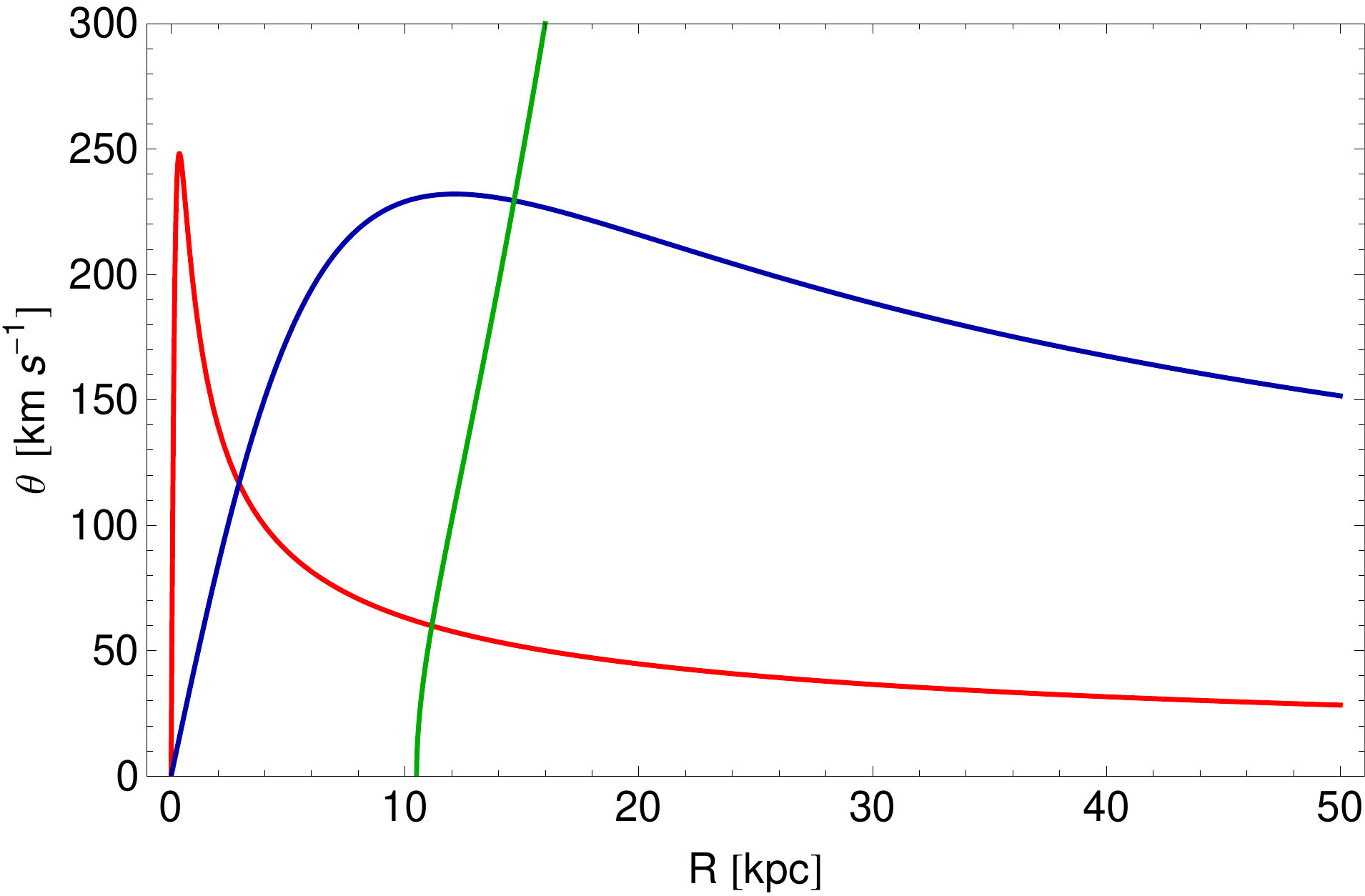}
\caption{Evolution of the three components of the circular velocity of the galactic model as a function of the distance $R$ from the galactic center. The red curve is the contribution from the spherical nucleus, the blue curve is the contribution from the disk, while the green curve corresponds to the perturbing terms.}
\label{rotcur}
\end{figure}

It is well known, that in axially symmetric systems like ours, the circular velocity in the galactic plane $z=0$,
\begin{equation}
\theta(R) = \sqrt{R\left|\frac{\partial V}{\partial R}\right|_{z=0}},
\label{cvel}
\end{equation}
is a very important physical quantity. The total potential $V(R,z)$ describing the properties of motion in our model consists of three components: a disk, a spherical nucleus and some additional perturbing terms. Therefore, the total circular velocity emerges by summing the contributions from the three distinct component; $\theta(R) = \theta_{\rm d}(R) + \theta_{\rm n}(R) + \theta_{\rm p}(R)$. In Fig. \ref{rotcur} we see the evolution of the three components of the circular velocity as a function of the distance $R$ from the galactic center. In particular, the red curve is the contribution from the spherical nucleus, the blue curve is the contribution from the disk, while the green curve corresponds to the perturbing terms. It is seen, that each contribution prevails in different distances form the galactic center. In particular, at small distances when $R \lesssim 3$ kpc, the contribution from the spherical nucleus dominates, while at mediocre distances, $3 < R < 10$ kpc, the disk contribution is the dominant factor. On the other hand, at large galactocentric distances $R > 10$ kpc, or in other words outside the main body of the galaxy, we observe that the circular velocity corresponding to the perturbing terms exhibits a very sharp increase being henceforth the prevailing term. Here, we should point out, that the particular values of the parameters entering the perturbation potential, which are given below, were chosen so that our model is as realistic as possible. This is true indeed, because the evolution of $\theta_{\rm p}(R)$ shown in Fig. \ref{rotcur} has a real physical meaning; it exists only in large galactocentric distances ($R > 10$ kpc), where the influence from nearby galaxies is strong, while inside the main galaxy is zeroed.

Taking into account that the total potential $V(R,z)$ is indeed axially symmetric, we know that in this case the $z$-component of the angular momentum $L_z$ is conserved. With this restriction, orbits can be described by means of the effective potential (e.g., [\citealp{3}])
\begin{equation}
V_{\rm eff}(R,z) = V(R,z) + \frac{L_z^2}{2R^2}.
\label{veff}
\end{equation}
The $L_z^2/(2R^2)$ term represents a centrifugal barrier and only orbits with small value of $L_z$ are allowed to pass near the axis of symmetry. Thus, the three-dimensional motion is effectively reduced to a two-dimensional motion in the meridional plane $(R,z)$, which rotates non-uniformly around the axis of symmetry according to
\begin{equation}
\dot{\phi} = \frac{L_z}{R^2},
\end{equation}
where of course the dot indicates derivative with respect to time.

The equations of motion on the meridional plane are
\begin{eqnarray}
\ddot{R} &=& - \frac{\partial V_{\rm eff}}{\partial R}, \nonumber \\
\ddot{z} &=& - \frac{\partial V_{\rm eff}}{\partial z},
\label{eqmot}
\end{eqnarray}
while the equations governing the evolution of a deviation vector $\delta{\bf w} = (\delta R, \delta z, \delta \dot{R}, \delta \dot{z})$ which joins the corresponding phase space points of two initially nearby orbits, needed for the calculation of the standard indicators of chaos, are given by the variational equations
\begin{eqnarray}
\dot{(\delta R)} &=& \delta \dot{R}, \nonumber \\
\dot{(\delta z)} &=& \delta \dot{z}, \nonumber \\
(\dot{\delta \dot{R}}) &=&
- \frac{\partial^2 V_{\rm eff}}{\partial R^2} \delta R
- \frac{\partial^2 V_{\rm eff}}{\partial R \partial z}\delta z,
\nonumber \\
(\dot{\delta \dot{z}}) &=&
- \frac{\partial^2 V_{\rm eff}}{\partial z \partial R} \delta R
- \frac{\partial^2 V_{\rm eff}}{\partial z^2}\delta z.
\label{vareq}
\end{eqnarray}

The corresponding Hamiltonian to the effective potential given in Eq. (\ref{veff}) can be written as
\begin{equation}
H = \frac{1}{2} \left(\dot{R}^2 + \dot{z}^2 \right) + V_{\rm eff}(R,z) = E,
\label{ham}
\end{equation}
where $\dot{R}$ and $\dot{z}$ are the momenta per unit mass conjugate to $R$ and $z$ respectively, while $E$ is the numerical value of the Hamiltonian (star's total energy), which is conserved. Therefore, all orbits are restricted to the area in the meridional plane satisfying $E \geq V_{\rm eff}$.

\begin{figure}
\includegraphics[width=\hsize]{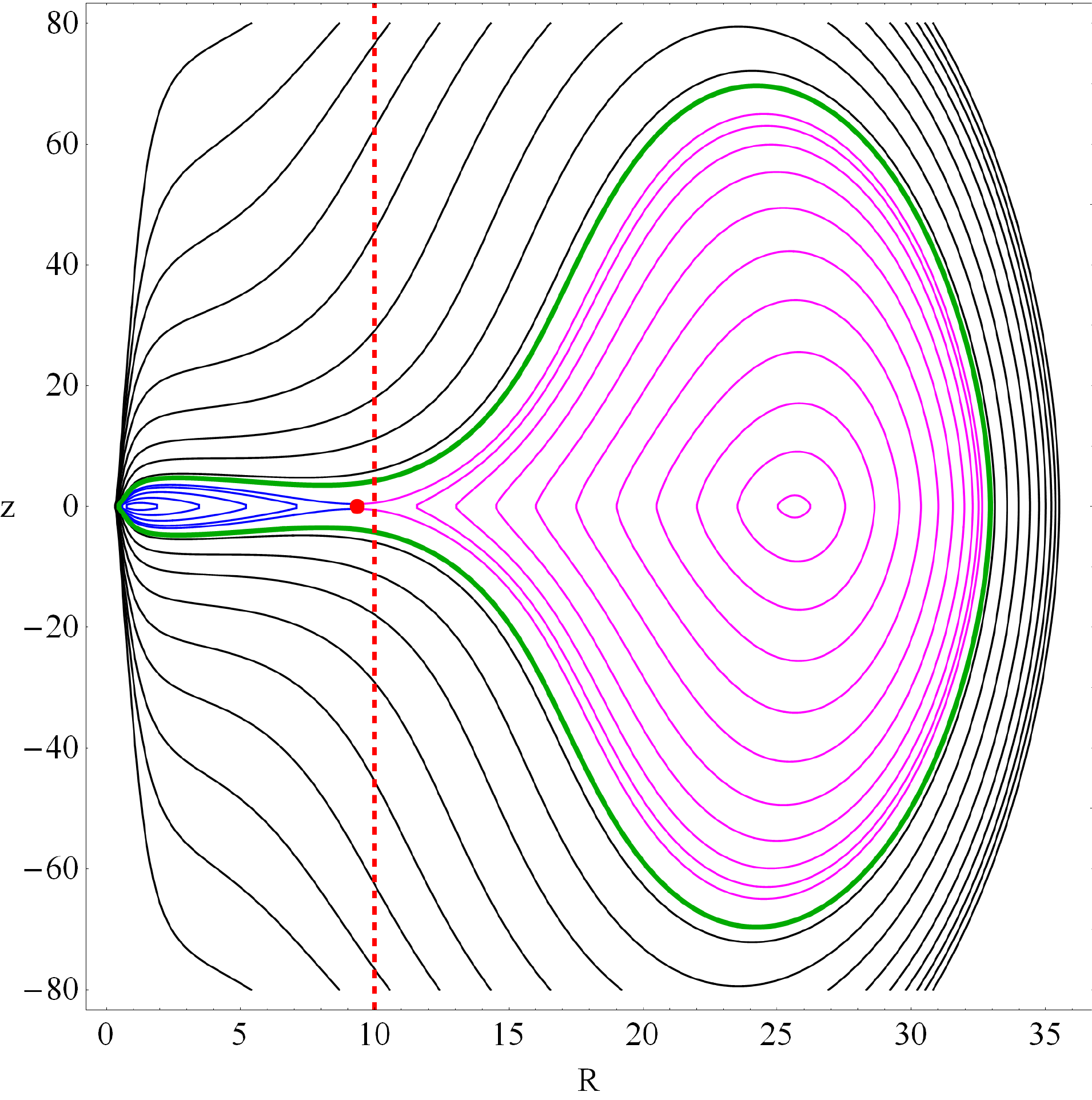}
\caption{A plot of the iso-potential curves for our galactic model when $M_{\rm n} = 400$ and $L_{\rm z} = 20$.}
\label{isopot}
\end{figure}

We use a system of galactic units where the unit of length is 1 kpc, the unit of velocity is 10 km s$^{-1}$, and $G=1$. Thus, the unit of mass results $2.325 \times 10^7 {\rm M}_\odot$, that of time is $0.9778 \times 10^8$ yr, the
unit of angular momentum (per unit mass) is 10 km$^{-1}$ kpc s$^{-1}$, and the unit of energy (per unit mass) is 100 km$^2$s$^{-2}$. We use throughout the paper the following values: $M_{\rm d} = 12000$, $b = 8$, $\alpha = 3$ and $h = 0.1$, $c_n = 0.25$, $k = -0.35$, $\lambda = 0.01$ and $\beta = 0.01$. The mass of the nucleus and the angular momentum on the other hand, are treated as parameters.

A plot of the iso-potential curves for our galactic model when $M_{\rm n} = 400$ and $L_{\rm z} = 20$ is presented in Fig. \ref{isopot}. The vertical, red, dashed line at $R = 10$ kpc marks the horizontal theoretical limit of the galaxy's main body. We observe, that there are three distinct types of contours: (i) blue iso-potential curves which are confined inside the main body of the galaxy, (ii) purple iso-potential curves which correspond to values of energy possessed only by distant stars and therefore, reach to large galactocentric distances ($R_{\ max} \gg$ 10 kpc) and (iii) black iso-potential curves that combine the two above-mentioned cases. In our research, we shall study the dynamical properties of stars at an energy level corresponding to the last case. In particular, we will use the value $E = -1100$ which remains constant throughout the paper. The iso-potential curve for this value of the energy is shown in green color in Fig. \ref{isopot}.

Moreover, in the same plot we see a red dot which indicates the main solution of the system
\begin{equation}
\left(\frac{\partial V_{\rm eff}(R,z)}{\partial R} = 0, \ \ \ \frac{\partial V_{\rm eff}(R,z)}{\partial z} = 0 \right).
\label{syssol}
\end{equation}
This particular point in the meridional $(R, z)$ plane is indeed a critical point, since it separates the first two families of iso-potential curves (contours around the main body of the galaxy and contours corresponding to distant orbits). Here we should notice, that this critical point is very close to the theoretical boundary ($R$ = 10 kpc) of the main body of the galaxy.

\section{Computational methods}
\label{cometh}

When studying the orbital structure of a dynamical system, knowing whether an orbit is regular or chaotic is an issue of significant importance. Over the years, several dynamical indicators have been developed in order to determine the nature of orbits. In our case, we chose to use the Smaller ALingment Index (SALI) method. The SALI [\citealp{33},\citealp{34}] is undoubtedly a very fast, reliable and effective tool, which is defined as
\begin{equation}
\rm SALI(t) \equiv min(d_-, d_+),
\label{sali}
\end{equation}
where $d_- \equiv \| {\vec{w_1}}(t) - {\vec{w_2}}(t) \|$ and $d_+ \equiv \| {\vec{w_1}}(t) + {\vec{w_2}}(t) \|$ are the alignments indices, while ${\vec{w_1}}(t)$ and ${\vec{w_2}}(t)$, are two deviations vectors which initially point in two random directions. For distinguishing between ordered and chaotic motion, all we have to do is to compute the SALI for a relatively short time interval of numerical integration $t_{max}$. More precisely, we track simultaneously the time-evolution of the main orbit itself as well as the two deviation vectors ${\vec{w_1}}(t)$ and ${\vec{w_2}}(t)$ in order to compute the SALI. The variational equations (\ref{vareq}), as usual, are used for the evolution and computation of the deviation vectors.

The time-evolution of SALI strongly depends on the nature of the computed orbit since when the orbit is regular the SALI exhibits small fluctuations around non zero values, while on the other hand, in the case of chaotic orbits the SALI after a small transient period it tends exponentially to zero approaching the limit of the accuracy of the computer $(10^{-16})$. Therefore, the particular time-evolution of the SALI allow us to distinguish fast and safely between regular and chaotic motion. Nevertheless, we have to define a specific numerical threshold value for determining the transition from regularity to chaos. After conducting extensive numerical experiments, integrating many sets of orbits, we conclude that a safe threshold value for the SALI taking into account the total integration time of $10^4$ time units is the value $10^{-8}$. In order to decide whether an orbit is regular or chaotic, one may use the usual method according to which we check after a certain and predefined time interval of numerical integration, if the value of SALI has become less than the established threshold value. Therefore, if SALI $\leq 10^{-8}$ the orbit is chaotic, while if SALI $ > 10^{-8}$ the orbit is regular. In Therefore, the distinction between regular and chaotic motion is clear and beyond any doubt when using the SALI method.

\begin{figure*}
\resizebox{\hsize}{!}{\includegraphics{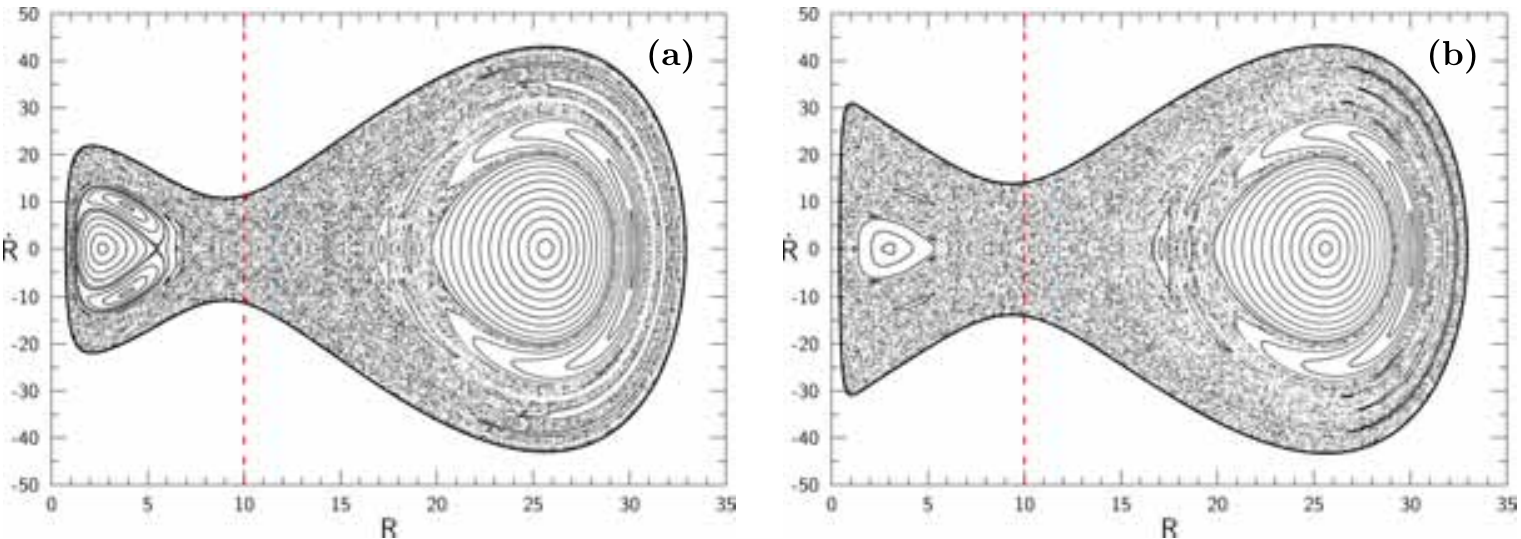}}
\caption{(a-b): The structure of the $(R,\dot{R})$ phase plane, when $L_{\rm z} = 20$ and (a-left): $M_{\rm n} = 50$ and (b-right): $M_{\rm n} = 400$.}
\label{PSSsMn}
\end{figure*}

\begin{figure*}
\resizebox{\hsize}{!}{\includegraphics{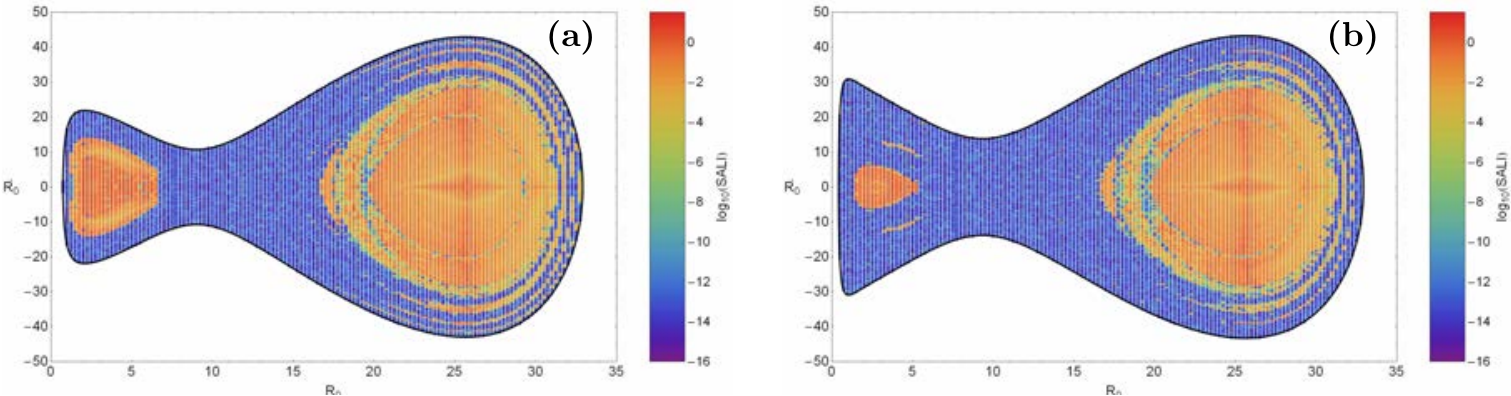}}
\caption{(a-b): Orbital structure of the $(R,\dot{R})$ phase plane, when $L_{\rm z} = 20$ and (a-left): $M_{\rm n} = 50$ and (b-right): $M_{\rm n} = 400$.}
\label{GridsMn}
\end{figure*}

For the study of our models, we need to define the sample of orbits whose properties (chaos or regularity) we will identify. The best method for this purpose, would have been to choose the sets of initial conditions of the orbits from a distribution function of the models. This, however, is not available so, we define, for each set of values of the parameters of the potential, a grid of initial conditions $(R_0, \dot{R_0})$ regularly distributed in the area allowed by the value of the energy. In each grid the step separation of the initial conditions along the $R$ and $\dot{R}$ axis was controlled in such a way that always there are at least 2 $\times$ $10^4$ orbits. For each initial condition, we integrated the equations of motion (\ref{eqmot}) as well as the variational equations (\ref{vareq}) using a double precision Bulirsch-Stoer FORTRAN algorithm [\citealp{28}] with a small time step of order of $10^{-2}$, which is sufficient enough for the desired accuracy of our computations (i.e. our results practically do not change by halving the time step). In all cases, the energy integral (Eq. (\ref{ham})) was conserved better than one part in $10^{-10}$, although for most orbits it was better than one part in $10^{-11}$.

\begin{figure*}
\resizebox{\hsize}{!}{\includegraphics{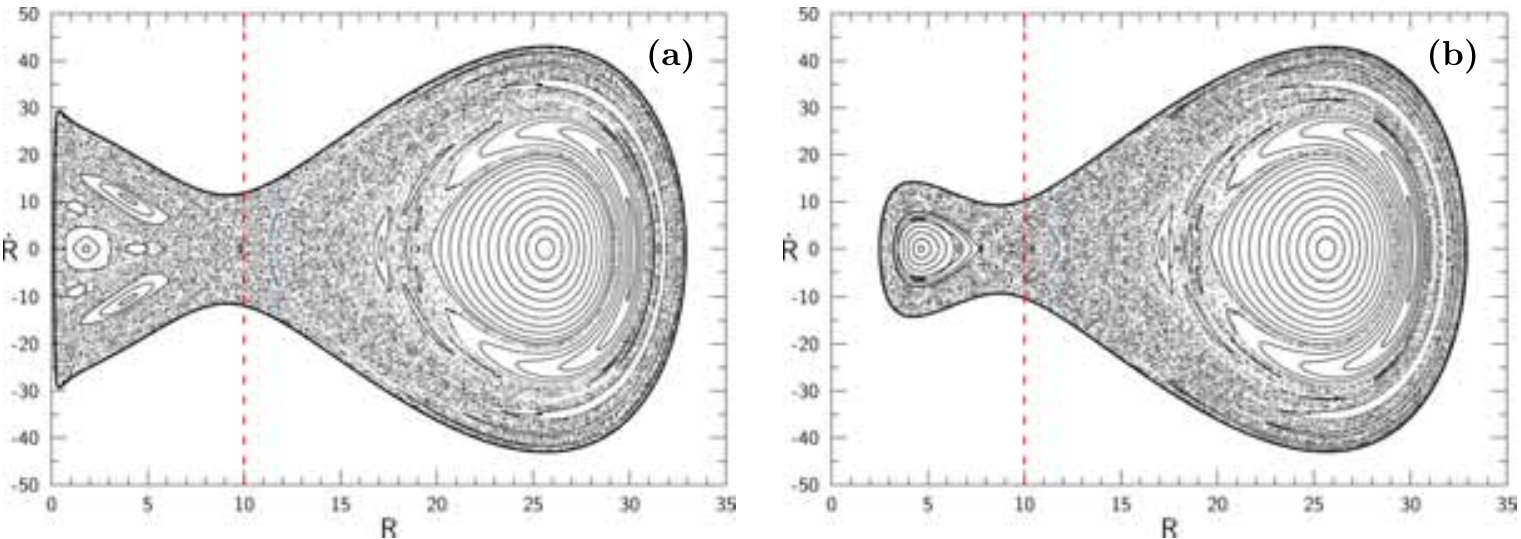}}
\caption{The structure of the $(R,\dot{R})$ phase plane, when $M_{\rm n} = 100$ and (a-left): $L_{\rm z} = 5$ and (b-right): $L_{\rm z} = 60$.}
\label{PSSsLz}
\end{figure*}

\begin{figure*}
\resizebox{\hsize}{!}{\includegraphics{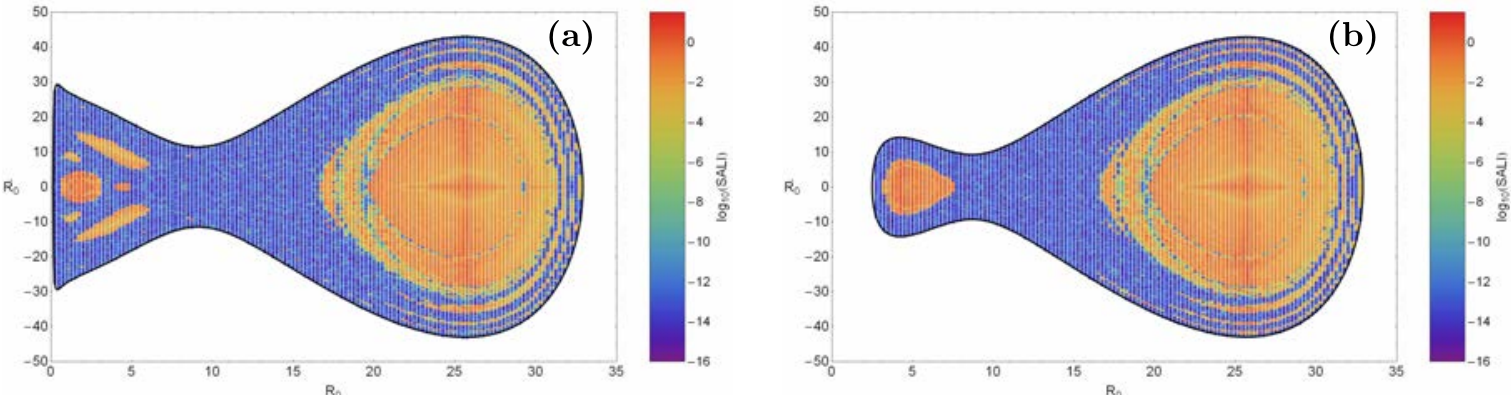}}
\caption{Orbital structure of the $(R,\dot{R})$ phase plane, when $M_{\rm n} = 100$ and (a-left): $L_{\rm z} = 5$ and (b-right): $L_{\rm z} = 60$.}
\label{GridsLz}
\end{figure*}

In our study, each orbit was integrated numerically for a time interval of $10^4$ time units ($10^{12}$ yr), which corresponds to a time span of the order of hundreds of orbital periods and about 100 Hubble times. The particular choice of the total integration time is an element of great importance, especially in the case of the so called ``sticky orbits" (i.e., chaotic orbits that behave as regular ones during long periods of time). A sticky orbit could be easily misclassified as regular by any chaos indicator\footnote{Generally, dynamical methods are broadly split into two types: (i) those based on the evolution of sets of deviation vectors in order to characterize an orbit and (ii) those based on the frequencies of the orbits which extract information about the nature of motion only through the basic orbital elements without the use of deviation vectors.}, if the total integration interval is too small, so that the orbit do not have enough time in order to reveal its true chaotic character. Thus, all the sets of orbits of a given grid were integrated, as we already said, for $10^4$ time units, thus avoiding sticky orbits with a stickiness at least of the order of 100 Hubble times. All the sticky orbits which do not show any signs of chaoticity for $10^4$ time units are counted as regular ones, since that vast sticky periods are completely out of scope of our research.

\section{Numerical results}
\label{numres}

In this section, we will complement the classical method of the $(R,\dot{R})$, $z = 0$, $\dot{z} > 0$ Poincar\'e Surface of Section (PSS) [\citealp{22}], in an attempt to visually distinguish the regular or chaotic nature of motion. We use the initial conditions mentioned in the previous section in order to build the respective PSSs, taking values inside the limiting curve defined by
\begin{equation}
\frac{1}{2} \dot{R}^2 + V_{\rm eff}(R,0) = E.
\label{zvc}
\end{equation}

Fig. \ref{PSSsMn}a depicts the phase plane when $L_{\rm z} = 20$ and $M_{\rm n} = 50$. One can observe a large unified chaotic sea, while there are also several islands of invariant curves corresponding to regular motion. In fact, there are two distinct areas of ordered orbits: (a) orbits that circulate inside only the main body of the galaxy and (b) orbits that live entirely outside the main body of the galaxy. In particular, there are three different types regarding the first kind of regular orbits: (i) 2:1 banana-type orbits corresponding to the invariant curves surrounding the central periodic point, (ii) box orbits that are located mainly outside of the 2:1 resonant orbits and (iii) 1:1 open linear orbits form the double set of elongated islands outside the 2:1 resonance. On the other hand, we see that all the chaotic orbits can reach large galactocentric distances ($R_{\rm max} \simeq 33$ kpc) and therefore, they are distant orbits. Throughout the paper the vertical, red, dashed line at $R = 10$ kpc marks the horizontal theoretical limit of the galaxy's main body, while the outermost black thick curve is the ZVC. In Fig. \ref{PSSsMn}b we present the structure of the PSS when $L_{\rm z} = 20$ and $M_{\rm n} = 400$, that is the case of a model with a more massive central nucleus. It is evident, that there are many differences with respect to Fig. \ref{PSSsMn}a which focus exclusively on the main body of the galaxy. The most visible differences is the growth of the region occupied by chaotic orbits, the decrease of the percentage of 1:1 resonant orbits and the increase in the allowed radial velocity $\dot{R}$ of the stars near the center of the galaxy. On the contrary, we could argue that the structure of the PSS when $R > 10$ kpc exhibits insignificant differences with respect to that shown in Fig. \ref{PSSsMn}a. Here we must note, that both PSSs shown in Fig. \ref{PSSsMn}(a-b) describe the nature of motion of test particles (stars) of low angular momentum with a variable mass of the nucleus.

Figs. \ref{GridsMn}a and \ref{GridsMn}b show grids of $(R_0, \dot{R_0})$ initial conditions of orbits that we have classified on the PSSs of Figs. \ref{PSSsMn}a and \ref{PSSsMn}b, respectively using the SALI method. In these grids, each point (orbit) is colored according to its $\rm log_{10}(SALI)$ value at the end of the integration. Here, the red color corresponds to regular orbits, the dark blue/purple color represents the chaotic orbits/regions, while all the intermediate colors between the two extreme ones, represent orbits having a small rate of local exponential divergence and/or orbits whose true chaotic character is revealed only after long integration time, for example, the so-called sticky orbits, that is, orbits that ``stick" on to quasi-periodic tori for long time intervals. Note, that the fraction of these peculiar orbits (which lie mainly around the borders of the islands of stability) is very small (only few per cent of the total amount of tested initial conditions) and therefore, can be discerned by eye in Fig. \ref{GridsMn}(a-b) only if one focuses on these particular regions. We should also point out the excellent agreement between the two methods (PSS and SALI grids) as far as the gross features are concerned, as well as the fact that the SALI can easily trace tiny regions of stability which correspond to small islands of invariant curves embedded in the chaotic sea which the PSS method has difficulties detecting them.

\begin{figure*}
\centering
\resizebox{0.90\hsize}{!}{\includegraphics{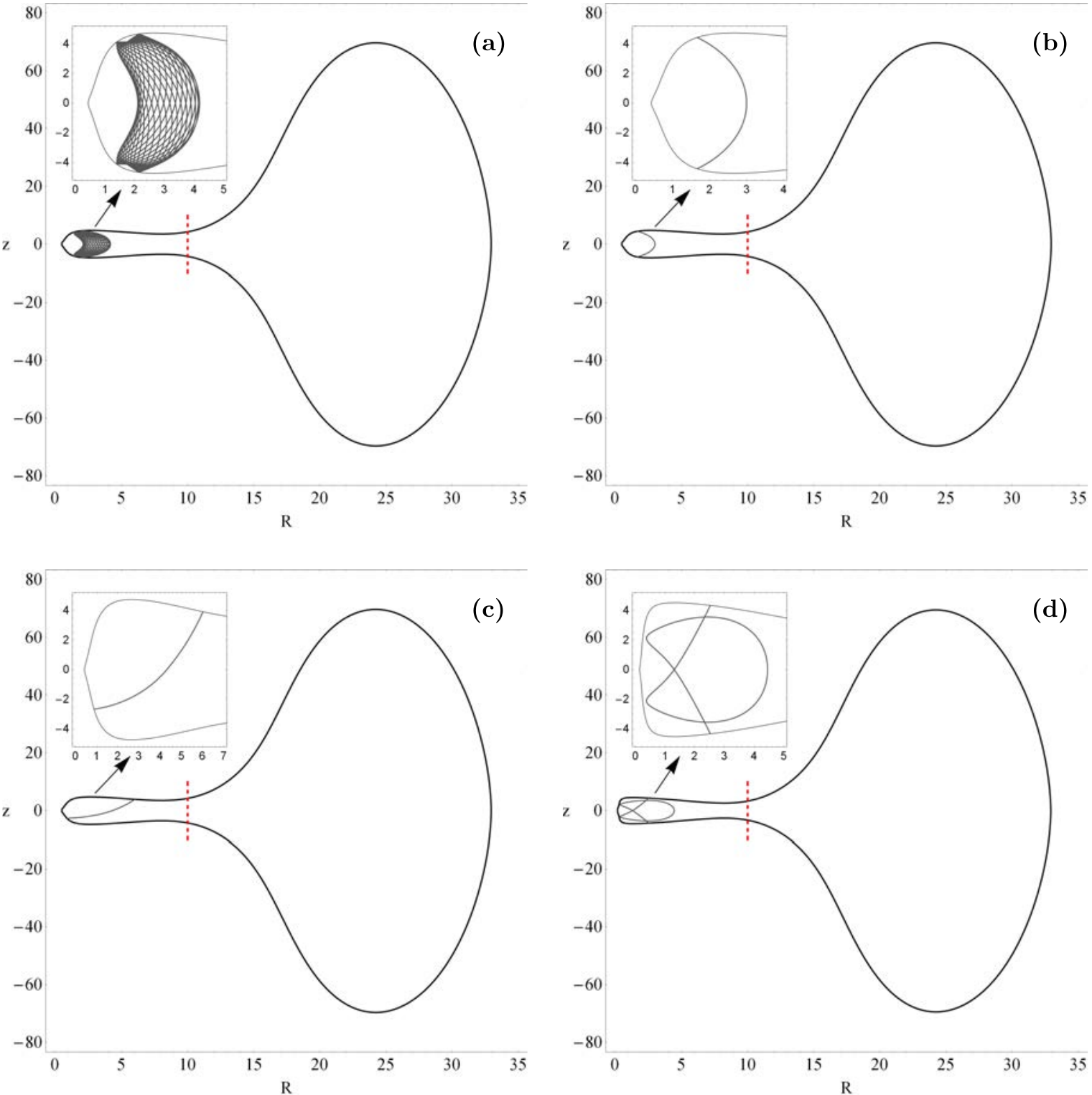}}
\caption{(a-d): Four characteristic examples depicting the basic types of regular orbits inside the main body of the galaxy.}
\label{OrbsGal}
\end{figure*}

\begin{figure*}
\centering
\resizebox{0.90\hsize}{!}{\includegraphics{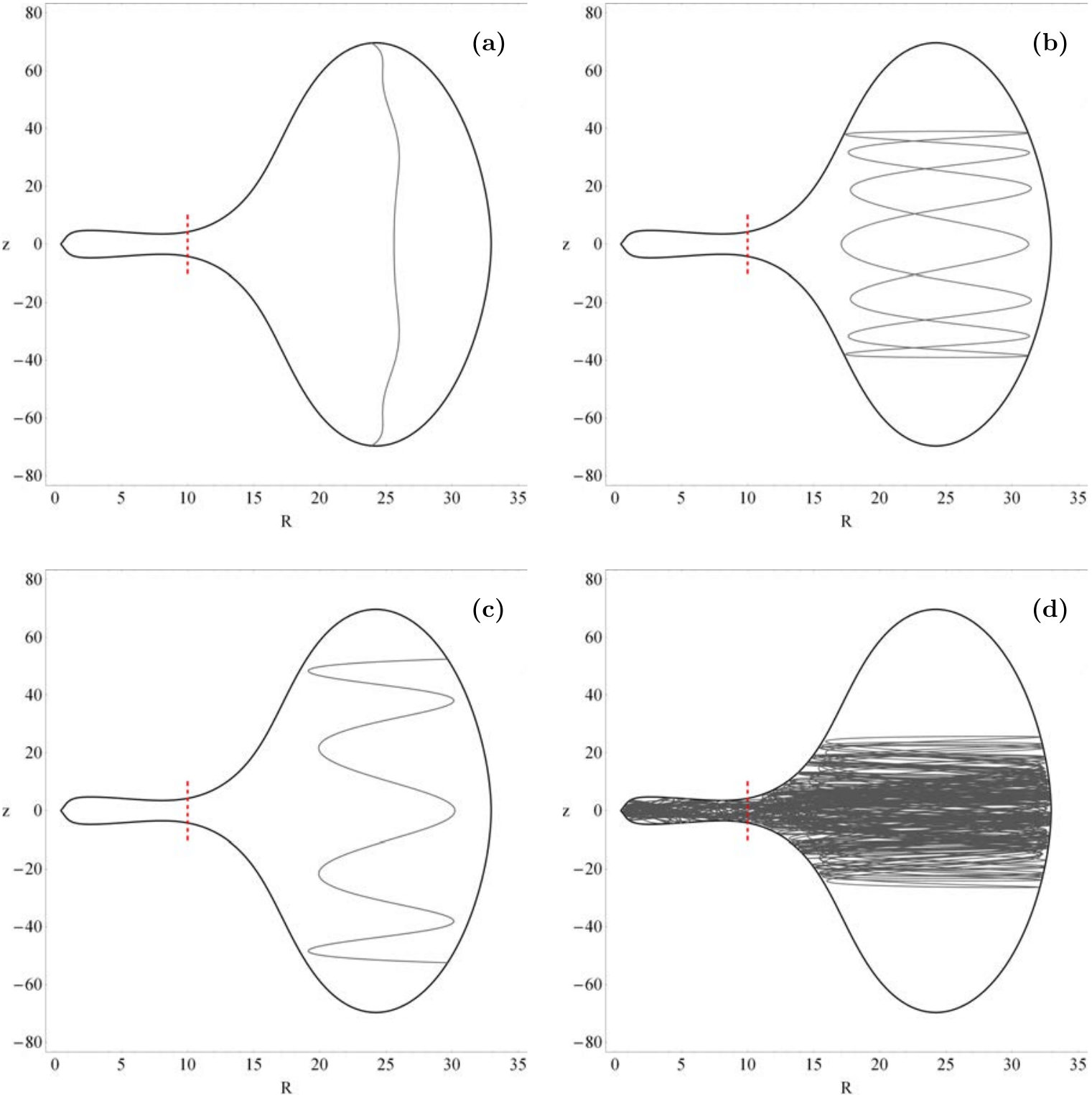}}
\caption{(a-d): Four typical examples of regular and chaotic distant orbits obtaining large galactocentric distances.}
\label{OrbsDis}
\end{figure*}

One of the most important parameters that influences significantly the orbital structure at the meridional plane is the angular momentum $L_z$. In this case, we let $L_z$ vary, while fixing $M_{\rm n} = 100$. Fig. \ref{PSSsLz}a presents the $(R,\dot{R})$ phase plane when $L_z = 5$ which correspond to motion of low angular momentum stars. We observe, that the overall structure of the PSS is very similar to the phase planes discussed previously. Again, there is a large unified chaotic sea which surrounds all the islands of stability. Inside the main body of galaxy there are four different types of regular orbits: (i) 2:1 banana-type orbits which correspond to the invariant curves surrounding the central periodic point in the corresponding PSS, (ii) box orbits that are situated mainly outside of the 2:1 resonant orbits, (iii) 1:1 open linear orbits form the double set of elongated islands in the PSS and (iv) 4:3 resonant orbits which correspond to the triple set of islands of invariant curves. Looking at Fig. \ref{PSSsLz}b, corresponding to $L_z = 60$, that is the case of high angular momentum stars, it is evident that the amount of chaos in the main body of the galaxy is smaller, while once more, the area where $R > 10$ kpc remains almost unaffected by the change of the value of the angular momentum. However, it is worth noticing that, at the highest angular momentum, the 1:1 resonance has been disappeared from the phase plane. From Fig. \ref{PSSsLz}(a-b) we can draw two important conclusions: (i) increasing $L_z$ causes a decreasing of the chaotic region inside the main body of the galaxy and (ii) the permissible area on the $(R,\dot{R})$ phase plane is reduced as we increase the value of $L_z$. Here we must point out, that both PSSs shown in Fig. \ref{PSSsLz}(a-b) describe the character of orbits of stars in galactic models with a mediocre mass of nucleus and a variable angular momentum. Figs. \ref{GridsLz}(a-b) show the grids of orbits corresponding to the PSSs of Figs. \ref{PSSsLz}(a-b) respectively. Once more, the excellent agreement between the two methods (PSS and SALI) is more than obvious.

\begin{figure}
\includegraphics[width=\hsize]{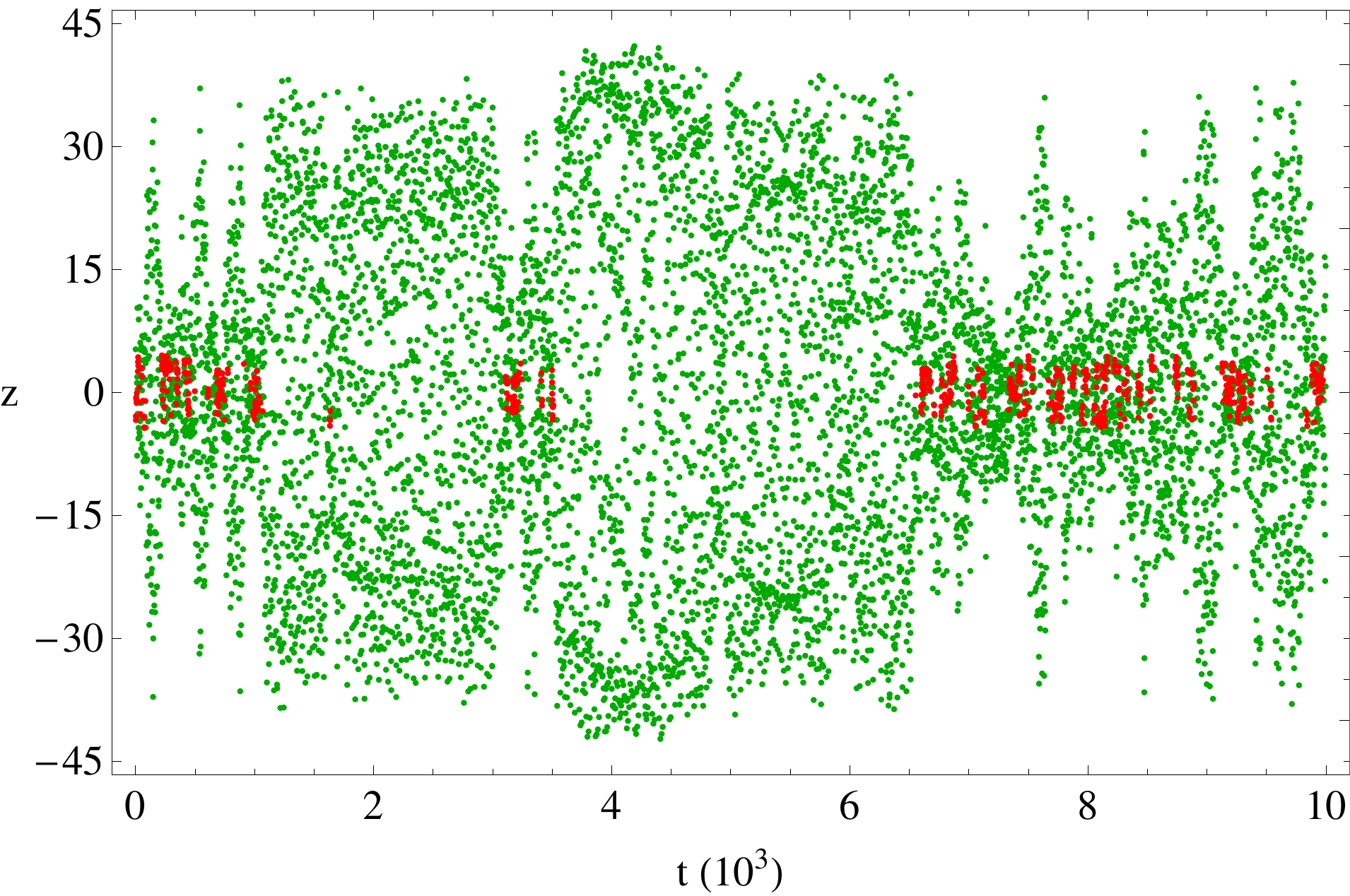}
\caption{Evolution of the $z$ height of a star following the chaotic orbit shown in Fig. \ref{OrbsDis}d. Red dots indicate time points when the star is inside the main body of the galaxy, while green dots correspond to time points where the star is moving into the galactic halo.}
\label{zevol}
\end{figure}

\begin{table}
   \caption{Types and initial conditions for the orbits shown in Figs. \ref{OrbsGal}(a-d) and \ref{OrbsDis}(a-d). In all
            cases, $z_0 = 0$, while $\dot{z_0}$ is found from the energy integral given by Eq. (\ref{ham}). $T_{\rm per}$ refers to the period of the parent periodic orbits.}
   \label{table}
   \setlength{\tabcolsep}{4.0pt}
   \begin{tabular}{@{}ccccc}
      \hline
      Figure & Type of orbit & $R_0$ & $\dot{R_0}$ & $T_{\rm per}$ \\
      \hline
      \ref{OrbsGal}a & box &  2.10000000 &  0.00000000 &           - \\
      \ref{OrbsGal}b & 2:1 &  2.99521125 &  0.00000000 &  1.30251998 \\
      \ref{OrbsGal}c & 1:1 &  4.30881125 & 11.50995880 &  1.24835455 \\
      \ref{OrbsGal}d & 4:3 &  4.45152018 &  0.00000000 &  3.94446962 \\
      \ref{OrbsDis}a & 2:1 & 25.60309520 &  0.00000000 & 11.25918720 \\
      \ref{OrbsDis}b & 7:1 & 17.08034875 &  0.00000000 &  9.73559785 \\
      \ref{OrbsDis}c & 8:1 & 30.20249838 &  0.00000000 & 10.64521257 \\
      \ref{OrbsDis}d & chaotic & 10.50000000 &  0.00000000 &       - \\
      \hline
   \end{tabular}
\end{table}

In Fig. \ref{OrbsGal}(a-d) we present four characteristic examples of the basic types of regular orbits that are encountered inside the main body of the galaxy. In all cases, the values of all the parameters are as in Fig. \ref{PSSsMn}b except for the orbit shown in Fig. \ref{OrbsGal}d, where the values of the parameters are as in Fig. \ref{PSSsLz}a. As expected, all orbits circulate close to the center of the galaxy and therefore, not only they do not reach large galactocentric distances but they do not even approach the boundary of the main galaxy's body. In fact, for all four orbits shown in Fig. \ref{OrbsGal}(a-b) we have that $\rm R_{max} \simeq 5$ kpc. In order to have a better view of these orbits we provide at the upper left part of each sub-panel a magnification of the area on the $(R,z)$ plane occupied by every orbit. The box orbit shown in Fig. \ref{OrbsGal}a was computed until $t = 100$ time units, while the parent periodic orbits were computed until one period has completed. The curve circumscribing each orbit is the limiting curve in the $(R,z)$ meridional plane defined as $V_{\rm eff}(R,z) = E$. In Table \ref{table} we provide the exact type and the initial conditions for each of the depicted orbits; for the resonant cases, the initial conditions and the period $T_{\rm per}$ correspond to the parent periodic orbit. Note that every resonance $n:m$ is expressed in such a way that $m$ is equal to the total number of islands of invariant curves produced in the $(R,\dot{R})$ phase plane by the corresponding orbit.

Things are quite different in Fig. \ref{OrbsDis}(a-d) where we observe four typical examples of distant orbits which reach to high galactocentric distances. The orbits presented in Fig. \ref{OrbsDis}(a-c) are resonant periodic orbits which exist outside the main galactic body, spent all their orbital time into the halo and therefore, exhibit large galactocentric distances. On the other hand, in Fig. \ref{OrbsDis}d we see a chaotic which stays near the galactic plane inside the main galaxy, while at large galactocentric distances it gains considerable height, since it consumes the vast majority of its orbital time into the halo. This particular but nevertheless interesting behavior of the orbit is dictated by the structure of the limiting curve (ZVC). It is of particular interest to note, that all the resonant periodic orbits shown in Fig. \ref{OrbsDis}(a-c) develop high departures from the galactic plane, in other words high values of $z$ coordinates and therefore, move on much deeper into the halo than the chaotic orbit. The initial conditions, the period and the exact type of the orbits are given in Table \ref{table}. All orbits were computed for a time interval equivalent to one period, except of the chaotic orbit shown in Fig. \ref{OrbsDis}d which was integrated for 200 time units.

In Fig. \ref{zevol} we present the evolution of the $z$ component of the chaotic orbit shown in Fig. \ref{OrbsDis}d for a time interval of $10^4$ time units. Red dots indicate the time points when the test particle (star) is inside the main body of the galaxy ($R < 10$ kpc), while green dots correspond to time points where the star is moving into the halo reaching large galactocentric distances. We observe, that the star moves randomly inside and outside the main galaxy. However, there are long time intervals of order of about 3000 time units, or even more, in which the star moves entirely outside the main body. Our numerical experiments indicate, that all stars moving in chaotic orbits in our dynamical system spent most of their orbital period (about 90\%) into the galactic halo, thus reaching large galactocentric distances and also high values of the $z$ coordinate.

\begin{figure*}
\resizebox{\hsize}{!}{\includegraphics{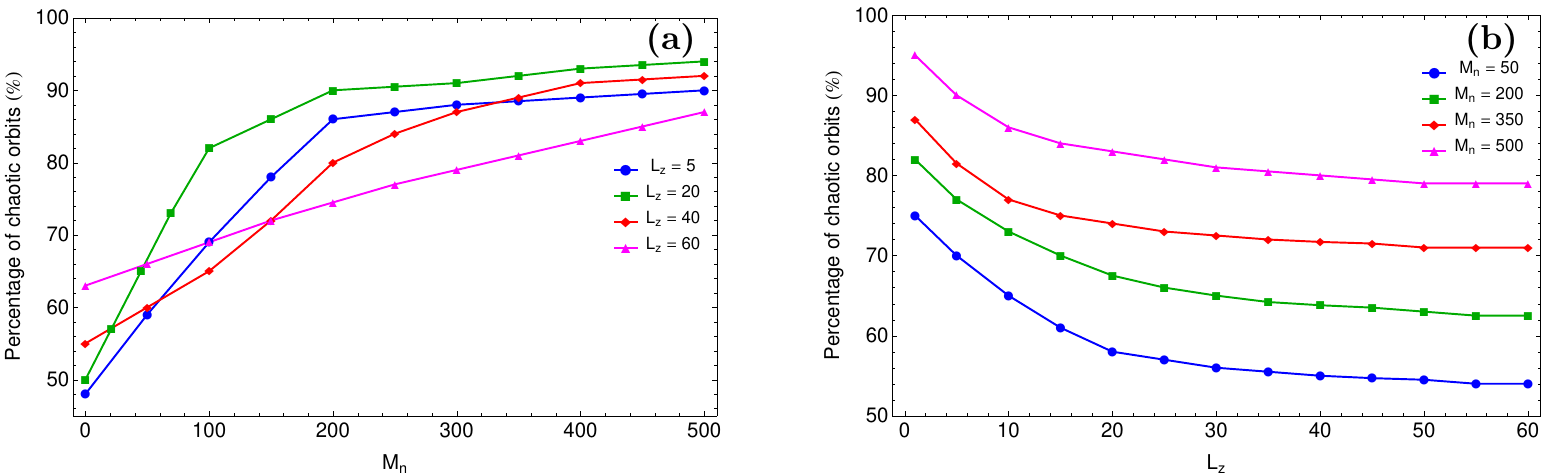}}
\caption{(a-b): A plot of the evolution of the percentage of the area covered by chaotic orbits in the $(R,\dot{R})$ phase plane as a function of (a-left): the mass of the nucleus for four values of the angular momentum and (b-right): the angular momentum for four values of the mass of the nucleus.}
\label{ChaosEvol}
\end{figure*}

We explained earlier, that in order to study the character of orbits in our galactic model, we integrated a set of initial conditions in each phase plane. Thus, calculating from all the sets of initial conditions the percentage of the chaotic orbits, we are able to follow how this fraction varies as a function of the mass of the nucleus and the angular momentum. The discrimination between the chaotic and regular orbits is that $\rm SALI < 10^{-8}$ for chaotic orbits, while $\rm SALI \geq 10^{-8}$ for regular ones. In the latter range, one does of course include the ``sticky" chaotic orbits.

To study how the mass of the nucleus $M_{\rm n}$ influences the level of chaos, we let it vary while fixing all the other parameters of our model. We chose $L_{\rm z} = 5, 20, 40$ and 60 for the angular momentum and integrated orbits in the meridional plane for the set $M_{\rm n} = \{0,50,100,...,500\}$. Fig. \ref{ChaosEvol}a shows the evolution of the percentage of the area covered by chaotic orbits in the $(R,\dot{R})$ phase planes as a function of the mass of the nucleus, for the four values of the angular momentum. We observe, that in general terms, the percentage of chaotic orbits increases as the mass of the nucleus increases in all four cases. However, the value of the angular momentum strongly affects the exact way of the evolution. It is clear form Fig. \ref{ChaosEvol}a that for low angular momentum stars the percentage of chaotic orbits increases rapidly when the mass of the nucleus is small enough. This is true until $M_{\rm n} \simeq 250$. We observe, that a further increase of the mass of the nucleus does not practically  affect the amount of chaos. On the other hand, the evolution of the chaotic percentage follow an entirely different path in the case of high angular momentum stars. For instance, when $L_{\rm z} = 60$ there is an almost linear increase of the chaotic percentage.

Now we proceed in order to investigate how the angular momentum $L_{\rm z}$ influences the amount of chaos in our galaxy model. Again, we let it vary while fixing all the other parameters of our galactic model, choosing $M_{\rm n} = 50, 200, 350$ and 500 as fiducial values for the mass of the nucleus, and integrating orbits in the meridional plane for the set $L_{\rm z} = \{1,5,10,15,...,60\}$. In Fig. \ref{ChaosEvol}b we present the evolution of the percentage of chaotic orbits in the $(R,\dot{R})$ phase planes as a function of the angular momentum, for the four values of the nucleus. It is evident, that the amount of chaos decreases with increasing angular momentum. With a more closer look at the diagram, we see that the more massive is the nucleus the more chaos is observed in the galaxy. For low values of angular momentum $(L_{\rm z} \leq 15)$ the chaotic percentage decreases sharply, while for larger values of $L_{\rm z}$ it remains almost the same. Nevertheless, we have to point out that the four trend lines corresponding to the evolution of the chaotic percentage of the four different values of $M_{\rm n}$ do not intersect, but they lined up one below the other following the reduction of the mass of the nucleus.

Before closing this section, we would like to make some comments regarding the existence of distant stars in the galactic halo. The mechanism responsible for the presence of all distant stars in the halo, is mainly the perturbation due to nearby galaxies. In fact, in a galaxy model without the extra perturbing terms, that is when $k = \lambda = 0$, there are no stars moving in remote, distant orbits. On the other hand, in a galaxy model where both the perturbing terms and the massive nucleus are present, the possibility of a star to have been ejected from the galactic plane is more than evident. Extensive numerical experiments, not presented here, indicate that the required time for a star to be ejected from the galactic plane thus moving in a distant orbit is about 50 time units (5 $\times$ $10^5$ yr), while the time needed for a distant star to return again inside the main bode of the galaxy is about 300 time units (3 $\times$ $10^{10}$ yr). On this basis, is becomes clear that the total number of distant stars in our galaxy model must be increasing.

Our galactic model suggests that a large portion of distant stars are indeed disk stars which have been scattered off the galactic plane into the halo moving in chaotic orbits. At this point, we would like to make clear that all these stars are not young O, B stars. In fact, the majority of these stars are very old. It is well known, that the central bulges in spiral galaxies contain old stars, while the disks of spirals contain a mixture of young and old stars. On the other hand, taking into account that the average age of B type stars is of order of several billion years, it seems more possible for those stars to have been formed into the halo. Another possible explanation justifying the presence of the small number of O, B stars at large z coordinates, could be the evolution in binary stellar systems with the subsequent mass exchange through a Roche-lobe flow [\citealp{23},\citealp{24}]. Finally, one should not overlook the possibility where stars are trapped in distant orbits upon their formation (see Figs. \ref{OrbsDis}(a-d)). These stars are old halo stars, probably RR-Lyrae type stars. This argument is strongly supported by data derived from observations, where faint RR-Lyrae type stars were discovered at large galactocentric distances up to 30 to 35 kpc from the galactic center [\citealp{30}-\citealp{32}].

\section{A semi-theoretical approach}
\label{semitheor}

\begin{figure}
\includegraphics[width=\hsize]{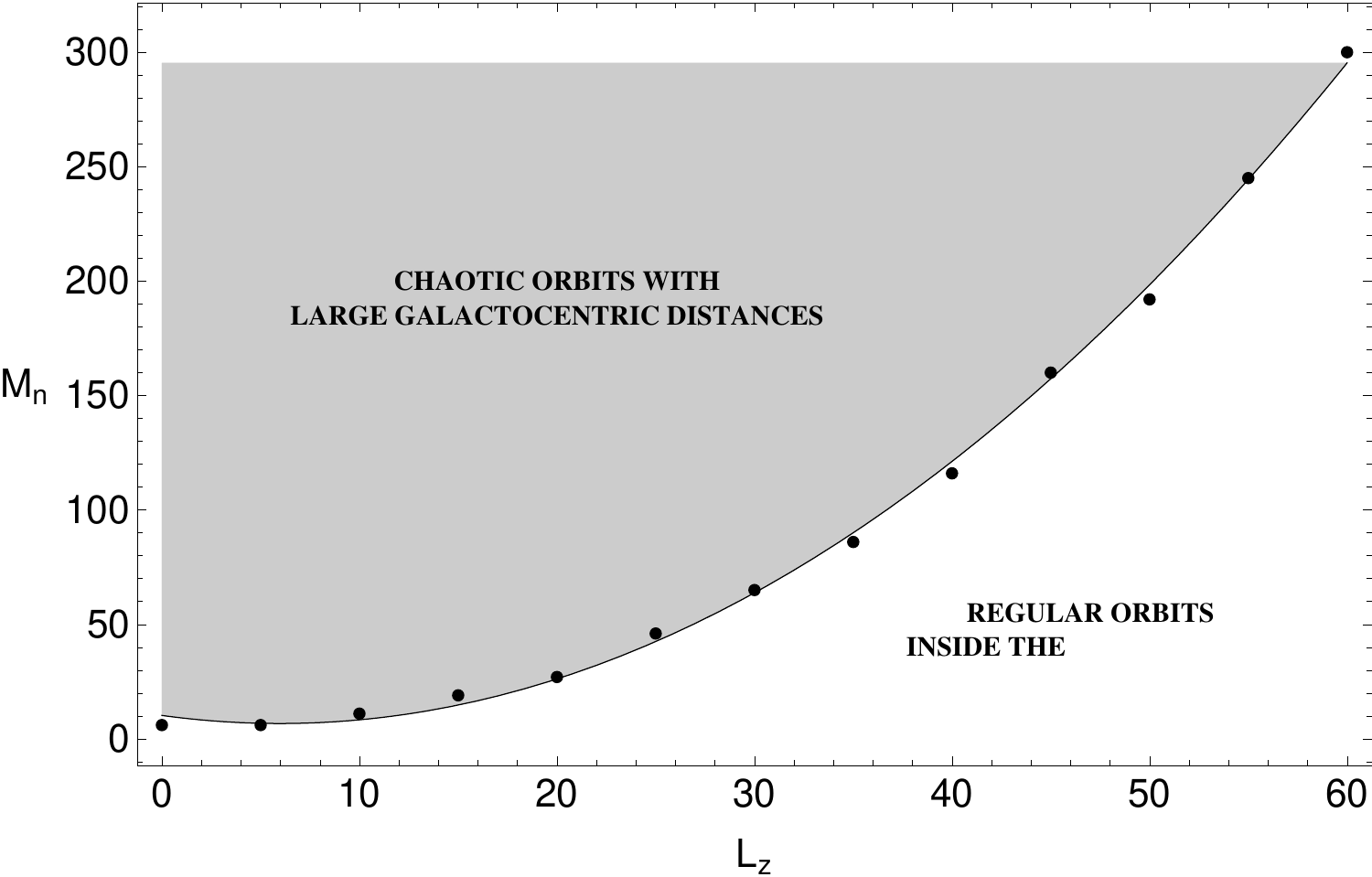}
\caption{Relationship between the critical value of the angular momentum $L_{\rm zc}$ and the mass of the nucleus $M_{\rm n}$. Details are given in the text.}
\label{LzcMn}
\end{figure}

In this section, we shall try to find if there is a relationship between the critical value of the angular momentum $L_{\rm zc}$ and the mass of the nucleus and if so, then try to explain and justify it using elementary semi-theoretical arguments. By the term ``critical value of the angular momentum" we refer to the maximum value of the angular momentum for which stars are scattered off the galactic plane into the halo thus displaying distant chaotic orbits, for a given value of the mass of the nucleus. A plot showing the relationship between $L_{\rm zc}$ and $M_{\rm n}$ is presented in Fig. \ref{LzcMn}. In order to obtain this correlation, we integrated numerically a large number of orbits. These orbits were started at $R_0 = 8.5$ kpc, $z_0 = 0$, with zero radial velocity $\dot{R_0}$, while the initial value of the vertical velocity $\dot{z_0}$ was always obtained from the energy integral (\ref{ham}). The numerically found results are indicated by dots, while the solid line which joins them is the best polynomial fit. In our case, there is a second degree polynomial dependence between the mass of the nucleus and the critical value of the angular momentum. Specifically, the best fitting curve is represented by the equation
\begin{equation}
M_{\rm n}(L_{\rm zc}) = 10.20881 - 1.18002 L_{\rm zc} + 0.09886 L_{\rm zc}^2,
\label{fitcur}
\end{equation}
which is indeed a second degree polynomial in $L_{\rm zc}$. Orbits with values of the parameters on the lower right part of the $(L_{\rm z},M_{\rm n})$ plane including the line correspond to regular orbits that are confined inside the main body of the galaxy ($R_{\rm max} \leq 10$ kpc), while orbits with values of the parameters on the upper left part of the same plane lead to chaotic orbits obtaining large galactocentric distances.

Now we are going to reproduce the form of Eq. (\ref{fitcur}) by combining some semi-theoretical arguments together with numerical evidence. We know from previous works, that the primary cause driving a star in large galactocentric orbit is the radial force $F_{\rm R}$ near the nucleus. On approaching the central nucleus there is a change in the star's angular momentum in the $R$ direction given by
\begin{equation}
m\Delta \upsilon_{\rm R} = \langle F_{\rm R} \rangle \Delta t,
\label{tre1}
\end{equation}
where $m$ is the mass of the star, $\langle F_{\rm R} \rangle$ is the average force acting along the $R$ direction near the nucleus and $\Delta t$ is the duration of the encounter. It was observed, that the test particle's (star) deflection far from the galactic center proceeds in each time cumulatively, a little more with each successive pass near the central nucleus and not with a single dramatic encounter. Let us assume that the star is ejected to a distant orbit after $n$ $(n > 1)$ passes, when the total change in the momentum in the $R$ direction is of order of $m \upsilon_{\rm \phi}$, where $\upsilon_{\rm \phi}$ is the tangential velocity of the star near the nucleus. Therefore we have
\begin{equation}
m \sum_{i=1}^n \Delta \upsilon_{\rm Ri} \approx \langle F_{\rm R} \rangle \sum_{i=1}^n \Delta t_i.
\label{tre2}
\end{equation}

If we set
\begin{eqnarray}
m &=& 1, \nonumber \\
\sum_{i=1}^n \Delta \upsilon_{\rm Ri} &=& - \upsilon_{\rm \phi} = - \frac{L_{\rm zc}}{R}, \nonumber \\
\sum_{i=1}^n \Delta t_i &=& T_{\rm c},
\label{tre3}
\end{eqnarray}
in Eq. (\ref{tre2}), we find
\begin{equation}
- \frac{L_{\rm zc}}{R T_{\rm c}} = \langle F_{\rm R} \rangle.
\label{tre4}
\end{equation}

Elementary numerical calculations reveal that, near the nucleus, where $R = R_0 < 1$ and $z \approx 0$ the $\langle F_{\rm R} \rangle$ force is repulsive and can be written in the following form
\begin{equation}
\langle F_{\rm R} \rangle \approx \frac{L_{\rm zc}^2}{R_0^3} - \frac{M_{\rm n} R_0}{\left(R_0^2 + c_{\rm n}^2\right)^{3/2}},
\label{tre5}
\end{equation}
because all the higher order terms are practically negligible near the nucleus. Inserting this value of $\langle F_{\rm R} \rangle$ in Eq. (\ref{tre4}) and after rearranging the order of the terms we find
\begin{equation}
M_{\rm n} \approx \left(\frac{L_{\rm zc}^2}{R_0^4} + \frac{L_{\rm zc}}{T_{\rm c} R_0^2}\right) \left(R_0^2 + c_{\rm n}^2 \right)^{3/2} \approx a_1 L_{\rm zc} + a_2 L_{\rm zc}^2,
\label{tre6}
\end{equation}
where $a_1$ and $a_2$ are constants. This is because the values of $R_0$ and $T_{\rm c}$ are about the same for the range of values of the mass of the nucleus and the angular momentum used in order to produce the diagram shown in Fig. \ref{LzcMn}. Eq. (\ref{tre6}) shows indeed a second degree polynomial relationship between $M_{\rm n}$ and $L_{\rm zc}$ but it is not complete yet. Numerical experiments indicate, that when $L_{\rm zc} \rightarrow 0$ there must be a minimum value $M_{\rm n0}$ of the mass of the nucleus in order to drive the star to a distant orbit (see near the origin at the plot in Fig. \ref{LzcMn}). Inserting this additional term in Eq. (\ref{tre6}), we finally obtain
\begin{equation}
M_{\rm n}(L_{\rm zc}) \approx M_{\rm n0} + a_1 L_{\rm zc} + a_2 L_{\rm zc}^2.
\label{tre7}
\end{equation}
Thus it is evident, that the form of Eq. (\ref{tre7}) is exactly the same as the numerically obtained relationship given by Eq. (\ref{fitcur}).

\section{Discussion and conclusions}
\label{disc}

The main objective of this research work, was the investigation of the character of orbits of stars in the meridional plane of an axially symmetric galactic gravitational model consisting of a disk, a dense spherical nucleus and some additional perturbing terms corresponding to interaction from nearby galaxies. Here, we have to point out that a large variety of dynamical models describing global motion in galaxies are also available in the literature (e.g., [\citealp{2},\citealp{13},\citealp{29},\citealp{35}]). Moreover, one must not forget the local galactic models which are mainly maid up of perturbed harmonic oscillators (e.g., [\citealp{8},\citealp{11},\citealp{16},\citealp{17},\citealp{37},\citealp{39},\citealp{40}]). The reader can find many illuminating information on dynamical galactic models in [\citealp{6}]. Here we have to point out, that there are only but a few similar research works on the nature of distant stars in galaxies. Therefore, we believe that our work makes a significant contribution to our so far knowledge of stars moving in large galactocentric distances under the perturbation of nearby galaxies.

In order to estimate the chaoticity of our models, we chose a dense grid of initial conditions in the $(R,\dot{R})$ phase plane, regularly distributed in the area allowed by the value of the energy and defined by the ZVC. All the samples of the orbits were integrated numerically and the regular or chaotic nature of each orbit had been determined by computing the SALI. In our galactic model three main types of orbits appear: (i) regular orbits that are confined inside the main body of the galaxy, (ii) distant regular orbits located at large galactocentric distances and (iii) chaotic orbits that spent the majority of their orbital time in large distances rather than inside the main galactic body and close to the central nucleus. One of the most interesting and surely novel aspects of this research was the study of the dynamical properties of the distant stars. Our results can be summarized as follows:

\begin{enumerate}
  \item The primary factor which is responsible for the existence of distant orbits is the presence of a dense and massive nucleus at the center of the galaxy, combined with a perturbation from nearby galaxies.
  \item The vast majority of the distant stars perform chaotic orbits, but there are also a significant amount of distant stars which display ordered motion.
  \item The vast majority of distant regular orbits are mainly $n:1$ resonant orbits, where $n > 2$, which go higher into the galactic halo thus, displaying large $z$ coordinates of order of about 50 kpc or even more.
  \item The mass of the nucleus, although spherically symmetric and therefore maintaining the axial symmetry of the whole galaxy, is one of the parameters which controls the percentage of chaos inside the main body of the galaxy. In particular, as the mass of the nucleus increases, the chaotic motion grows in percentage. Similar results regarding the effect of the mass of the nucleus had been found in a recent work [\citealp{38}].
  \item The value of the angular momentum of the orbits also influences the level of chaos inside the main galaxy. The percentage of the chaotic orbits decreases with increasing angular momentum.
  \item It seems that both the mass of the nucleus and the angular momentum have a relatively small radius of influence which is almost confined to the boundaries of the main galaxy. Beyond this limit ($R_{\rm max} \simeq 10$ kpc) the structure of the phase plane remains almost the same and the differences due to $M_{\rm n}$ or $L_{\rm z}$, if any, are negligible.
  \item The mass of the nucleus and the critical value of the angular momentum of the distant stars are linked through a second order polynomial relationship. The particular polynomial law had been reproduced and therefore justified using elementary semi-theoretical arguments.
\end{enumerate}

According to current data, it is evident that the majority of distant stars are in fact old stars. Therefore, the presence of young O and B stars in the galactic halo is, by all means, a very interesting issue that needs to be explored and clarified. So far, there are two possible scenarios that can, in a way, explain the young O and B distant stars observed in large galactocentric distances. According to the first one, these stars might have been formed initially into the halo. On the other hand, the second scenario suggests the evolution in binary stellar systems with the consequent mass exchange through a Roche-lobe flow. Taking all facts into account we strongly believe, that more observational data are needed, in order to be able to determine, once and for all, not only the origin but also the mysterious nature of the young distant O and B stars.

\section*{Acknowledgments}

I would like to express my warmest thanks to the two anonymous referees for the careful reading of the manuscript and for all the aptly suggestions and comments which improved both the quality and the clarity of the paper.

\end{document}